\documentstyle{elsart}

\def\go{\mathrel{\raise.3ex\hbox{$>$}\mkern-14mu\lower0.6ex\hbox{$\sim$}}}
\def\lo{\mathrel{\raise.3ex\hbox{$<$}\mkern-14mu\lower0.6ex\hbox{$\sim$}}}

\begin{document}

\begin{frontmatter}
\title{Smoothed Particle Hydrodynamics Calculations of Stellar Interactions}

\author[AUT1]{Frederic A.~Rasio}
and
\author[AUT2]{James C.~Lombardi, Jr.}
\address[AUT1]{Department of Physics, M.I.T., Cambridge, MA 02139, USA}
\address[AUT2]{Department of Physics, Vassar College, Poughkeepsie, NY 12604, USA}

\begin{abstract}
Smoothed Particle Hydrodynamics is a multidimensional Lagrangian method 
of numerical hydrodynamics that has been used to tackle a wide variety of
problems in astrophysics. Here we develop the basic equations of the
SPH scheme, and we discuss some of its numerical properties and limitations.
As an illustration of typical astrophysical applications, we discuss recent
calculations of stellar interactions, including collisions between main sequence
stars and the coalescence of compact binaries.
\end{abstract}

\end{frontmatter}

\section{Smoothed Particle Hydrodynamics}

Smoothed Particle Hydrodynamics (SPH) is a Lagrangian method that was 
introduced specifically to simulate 
self-gravitating fluids moving freely in three dimensions.
The key idea of SPH is to calculate pressure gradient forces by kernel
estimation, directly from the particle positions, rather than by finite
differencing on a grid, as in older particle methods such
as PIC. SPH was first introduced by 
Lucy (1977) and Gingold \& Monaghan (1977), who used it to study dynamical 
fission instabilities in rapidly rotating stars. Since
then, a wide variety of astrophysical fluid dynamics problems have been
tackled using SPH (see Monaghan 1992 for an overview). 
In addition to the stellar interaction problems described in \S2,
these have included
planet and star formation (Nelson et al.\ 1998; Burkert et al.\ 1997), 
supernova explosions (Herant \& Benz 1992; Garcia-Senz et al.\ 1998), 
large-scale cosmological structure formation (Katz et al.\ 1996;
 Shapiro et al.\ 1996), and
galaxy formation (Katz 1992; Steinmetz 1996).

\subsection{SPH from a Variational Principle}

A straightforward derivation of the basic SPH equations
can be obtained from a Lagrangian formulation of hydrodynamics
(Gingold \& Monaghan 1982). Consider for
simplicity the adiabatic evolution of an ideal fluid with equation of state
\begin{equation}
p=A\rho^\gamma, \label{eos}
\end{equation}
where $p$ is the pressure, $\rho$ is the density, $\gamma$ is the adiabatic
exponent, and $A$ (assumed here
to be constant in space and time) is related to the specific
entropy ($s \propto \ln A$). The Euler equations of motion,
\begin{equation}
{{\d}\vec{v}\over \d t}=
   {\partial\vec{v}\over\partial t}+(\vec{v}\cdot\nabla)\vec{v}
     =-{1\over\rho}\nabla p,
\end{equation}
can be derived from a variational principle with the Lagrangian
\begin{equation}
L=\int\left\{{1\over2}v^2 - u[\rho(\vec{r})]\right\}\,\rho\,{\d}^3x.
\end{equation}
Here 
$u[\rho]=p/[(\gamma-1)\rho]=A\rho^{\gamma-1}/(\gamma-1)$ is the specific internal 
energy of the fluid. 

The basic idea in SPH is to use the discrete representation 
\begin{equation}
L_{SPH}=\sum_{i=1}^N\, m_i\left[{1\over2}v_i^2 -u(\rho_i)\right] \label{lsph}
\end{equation}
for the Lagrangian, where the sum is over a large but discrete number of small
fluid elements, or ``particles,'' covering the volume of the fluid. Here $m_i$ 
is the mass and $\vec{v}_i$ is the velocity of the particle with position 
$\vec{r}_i$.  For expression~(\ref{lsph}) to become the Lagrangian of a system 
with a finite number $N$ of
degrees of freedom, we need a prescription to compute the density 
$\rho_i$ at the
position of any given particle $i$, as a function of the masses and positions of
neighboring particles.

In SPH, the density at any position is typically calculated as the
local average
\begin{equation}
\rho(\vec{r})=\sum_j m_j W(\vec{r}-\vec{r}_j;\, h),\label{rho}
\end{equation}
where $W(\vec{x};\,h)$ is an interpolation, or smoothing, kernel of
width $\sim h$.  Necessary constraints on the kernel
$W(\vec{x};\,h)$ are that (i) it integrates to unity (consequently the
integral of eq. (\ref{rho}) over all space automatically gives the
total mass of the system), and (ii) it approaches the Dirac delta
function $\delta(\vec{x})$ in the limit $h\rightarrow 0$.

Equation (\ref{rho}) gives, in particular, the density in the vicinity of 
particle $i$ as $\rho_i=\rho(\vec{r}_i)$, and we can now obtain the equations of motion
for all the particles. Deriving the Euler-Lagrange equations from $L_{SPH}$ we get
\begin{equation}
{{\d}\vec{v}_i\over \d t}=-\sum_j m_j \,\left({p_i\over\rho_i^2}
                   +{p_j\over\rho_j^2}\right)\, \nabla_i W_{ij}, \label{simple}
\end{equation}
where $W_{ij}= W(\vec{r}_i-\vec{r}_j;\, h)$ and we have assumed that
the form of $W$ is such that $W_{ij}=W_{ji}$.
The expression on the
right-hand side of eq.~(\ref{simple}) is a sum over neighboring particles (within a 
distance $\sim h$ of $\vec{r}_i$) 
representing a discrete approximation to the pressure gradient
force $[-(1/\rho)\nabla p]_i$ acting on particle $i$.

The following energy and momentum conservation laws are satisfied {\em exactly\/}
by the simple SPH equations of motion given above
\begin{equation}
{\d\over \d t}\left(\sum_{i=1}^N m_i \vec{v}_i \right) =0,
\end{equation}
and
\begin{equation}
{\d\over \d t}\left(\sum_{i=1}^N m_i\, [{1\over2}v_i^2 +u_i]\right) =0,
\end{equation}
where $u_i=p_i/[(\gamma-1)\rho_i]$.
Note that energy and momentum conservation in this simple version of
SPH is independent of the number of particles $N$.

Typically, a full implementation of SPH for astrophysical problems will add
to eq.~(\ref{simple}) a treatment of self-gravity (e.g., using one of the many 
grid-based or tree-based algorithms developed for N-body simulations) and an
artificial viscosity term to allow for entropy production in shocks.  
In addition, we have assumed here that the smoothing length
$h$ is constant in time and the same for all particles. In practice, individual
and time-varying smoothing lengths $h_i(t)$ are almost always used, so that the
local spatial resolution can be adapted to the (time-varying) density of SPH particles
(see Nelson \& Papaloizou 1994 for a rigorous derivation of the equations of 
motion from a variational principle in this case). Other derivations of the SPH
equations, based on the application of smoothing operators to the fluid equations
(and without the use of a variational principle), are also possible (see, e.g.,
Hernquist \& Katz 1989).

\subsection{Basic SPH Equations}

In this section, we summarize the basic equations for various forms
of the SPH scheme currently in use,
incorporating gravity, artificial viscosity, and individual smoothing lengths.

\subsubsection{Density and Pressure}

The SPH estimate of the fluid density at $\vec{r}_i$ is
calculated as $\rho_i=\sum_j m_j W_{ij}$ [cf.\ eq.~(\ref{rho})].  
Many recent implementations of
SPH use a form for $W_{ij}$ proposed by Hernquist \& Katz (1989),
\begin{equation}
W_{ij}={1\over2}\left[W(|\vec{r}_i-\vec{r}_j|;\,h_i)+W(|\vec{r}_i-
\vec{r}_j|;\,h_j)\right].
\end{equation}
This choice guarantees symmetric weights $W_{ij}=W_{ji}$ even between
particles $i$ and $j$ with different smoothing lengths.
For the interpolation kernel $W(r;\,h)$, the cubic spline
\begin{equation}
W(r;\,h)={1\over\pi h^3}
\cases{1-{3\over2}\left({r\over h}\right)^2
      +{3\over4}\left({r\over h}\right)^3,        & $0\le{r\over h}<1$,\cr 
{1\over4}\left[2-\left({r\over h}\right)\right]^3,& $1\le{r\over h}<2$,\cr
      0,                                          & ${r\over h}\ge2$,\cr} \label{WML}
\end{equation}
(Monaghan \& Lattanzio 1985) is a common choice.
Eq.~(\ref{WML}) is sometimes called a ``second-order accurate'' kernel.
Indeed, when the true density $\rho(\vec{r})$ of the fluid is represented by an
appropriate distribution of particle positions, masses, and smoothing
lengths, one can show that $\rho_i=\rho(\vec{r}_i)+O(h_i^2)$
(see, e.g., Monaghan 1985).

Depending on which thermodynamic evolution equation is integrated [see
eqs.~(\ref{udot}) and~(\ref{adot}) below], particle $i$ also
carries either the parameter $u_i$, the internal energy per
unit mass in the fluid at $\vec{r}_i$, or $A_i$, the entropic variable,
a function of the specific entropy in the fluid at $\vec{r}_i$.
Although arbitrary
equations of state can be implemented in SPH, here, for simplicity,
we consider only polytropic equations of state.
The pressure $p_i$ at $\vec{r}_i$ is therefore related to the density by
\begin{equation}
p_i=(\gamma-1)\,\rho_i\, u_i,
\end{equation}
or
\begin{equation}
p_i=A_i\,\rho_i^\gamma.
\end{equation}
The speed of sound in the fluid at $\vec{r}_i$ is $c_i=(\gamma
p_i/\rho_i)^{1/2}$.  

\subsubsection{Dynamical Equations and Gravity}

Particle positions are updated either by
\begin{equation}
{{\d}\vec{r}_i \over\d t}= \vec{v}_i, \label{rdot}
\end{equation}
or the more general XSPH method
\begin{equation}
{{\d}\vec{r}_i\over \d t} = \vec{v}_i+\epsilon \sum_j m_j{\vec{v}_j-\vec{v}_i\over
\rho_{ij}}W_{ij} \label{XSPH}
\end{equation}
where $\rho_{ij}=(\rho_i+\rho_j)/2$ and $\epsilon$ is a constant
parameter in the range $0 < \epsilon < 1$ (Monaghan 1989).  Eq.~(\ref{XSPH}), 
in contrast to eq.~(\ref{rdot}), changes particle
positions at a rate closer to the local smoothed velocity.  The XSPH
method was originally proposed as a way to minimize spurious interparticle
penetration across the interface of two colliding fluid streams.

Generalizing equation (\ref{simple}) to account for gravitational forces 
and artificial viscosity (hereafter AV), 
the velocity of particle $i$ is updated according to
\begin{equation}
           {{\d} \vec{v}_i\over \d t} = \vec{a}^{(Grav)}_i+\vec{a}^{(SPH)}_i
\end{equation}
where $\vec{a}^{(Grav)}_i$ is the gravitational acceleration and
\begin{equation}
\vec{a}^{(SPH)}_i=-\sum_j m_j \left[\left({p_i\over\rho_i^2}+
    {p_j\over\rho_j^2}\right)+\Pi_{ij}\right]{\bf \nabla}_i W_{ij}.
    \label{fsph}
\end{equation}
Various forms for the AV term $\Pi_{ij}$ are discussed below.
The AV ensures that correct jump
conditions are satisfied across (smoothed) shock fronts, while the rest of
equation~(\ref{fsph}) represents one of many possible SPH-estimators
for the acceleration due to the local pressure gradient (see, e.g.,
Monaghan 1985).

To provide reasonable accuracy, an SPH code must solve the equations of motion 
of a large number of particles (typically $N>>1000$). This rules out a
direct summation method for calculating the gravitational field of the
system, unless special purpose hardware such as the GRAPE is used
(Steinmetz 1996; Klessen 1997).  In most implementations of SPH,
particle-mesh algorithms (Evrard 1988; Rasio \& Shapiro
1992; Couchman et al.\ 1995) 
or tree-based algorithms (Hernquist \& Katz 1989; Dave et al.\ 1997) 
are used to calculate the gravitational accelerations $\vec{a}^{(Grav)}_i$.  
Tree-based algorithms perform better for problems involving large
dynamic ranges in density, such as star formation and large-scale cosmological
simulations. For stellar interaction problems like those described in
\S 2, density contrasts rarely exceed a factor $\sim10^2-10^3$ and in those
cases grid-based algorithms and direct solvers are generally faster.
Tree-based and grid-based algorithms are also used to calculate lists
of nearest neighbors for each particle exactly as in gravitational
$N$-body simulations.

\subsubsection{Artificial Viscosity}

For the AV, a symmetrized version of the form
proposed by Monaghan (1989) is often adopted,
\begin{equation}
\Pi_{ij}={-\alpha\mu_{ij}c_{ij}+\beta\mu_{ij}^2\over\rho_{ij}},
\label{pi}
\end{equation}
where $\alpha$ and $\beta$ are constant parameters,
$c_{ij}=(c_i+c_j)/2$, and
\begin{equation}
\mu_{ij}=\cases{ {\left(\vec{v}_i-\vec{v}_j\right)\cdot(\vec{r}_i-\vec{r}_j)\over
h_{ij}\left(|\vec{r}_i -\vec{r}_j|^2/h_{ij}^2+\eta^2\right)}& if $(\vec{v}_i-
\vec{v}_j)\cdot(\vec{r}_i-\vec{r}_j)<0$\cr
	     0& if $(\vec{v}_i-\vec{v}_j)\cdot(\vec{r}_i-
             \vec{r}_j)\ge0$\cr}
		\label{mu}
\end{equation}
with $h_{ij}=(h_i+h_j)/2$.  
This form represents a combination of a bulk
viscosity (linear in $\mu_{ij}$) and a von~Neumann-Richtmyer 
viscosity (quadratic in $\mu_{ij}$).  The von~Neumann-Richtmyer
viscosity was initially introduced to suppress particle
interpenetration in the presence of strong shocks.  Eq.~(\ref{pi})
provides a good treatment of shocks when $\alpha\approx1$,
$\beta\approx2$ and $\eta^2\sim10^{-2}$ (Monaghan 1989; Hernquist \&
Katz 1989).

A well known problem with the classical AV of eq.~(\ref{pi})
is that it can generate
large amounts of spurious shear viscosity.  For this reason, Hernquist \& Katz (1989)
introduced another form for the AV:
\begin{equation}
\Pi_{ij}=\cases{ {q_i\over\rho_{i}^2}+{q_j\over\rho_{j}^2}& if $(
\vec{v}_i-\vec{v}_j)\cdot(\vec{r}_i-\vec{r}_j)<0$\cr
	     0& if $(\vec{v}_i-\vec{v}_j)\cdot(\vec{r}_i-
             \vec{r}_j)\ge0$\cr},
		\label{pi2}
\end{equation}
where
\begin{equation}
q_i=\cases{ \alpha \rho_i c_i h_i |{\bf \nabla}\cdot \vec{v}|_i+
	       \beta \rho_i h_i^2 |{\bf \nabla}\cdot \vec{v}|_i^2
		 & if $\left({\bf \nabla}\cdot \vec{v}\right)_i<0$\cr
             	0& if $\left({\bf \nabla}\cdot \vec{v}\right)_i\ge0$\cr}
			\label{q}
\end{equation}
and
\begin{equation}
({\bf \nabla}\cdot \vec{v})_i={1 \over \rho_i}\sum_j m_j
	(\vec{v}_j-\vec{v}_i)\cdot{\bf \nabla}_i W_{ij}. \label{divv}
\end{equation}
Although this form provides a slightly less accurate description of 
shocks than equation (\ref{pi}), it does
exhibit less shear viscosity.

More recently, Balsara (1995) has proposed the AV
\begin{equation}
\Pi_{ij}=
\left({p_i\over\rho_i^2}+{p_j\over\rho_j^2}\right)
	\left(-\alpha \mu_{ij} + \beta \mu_{ij}^2\right),
	\label{piDB}
\end{equation}
where
\begin{equation}
\mu_{ij}=\cases{ {(\vec{v}_i-\vec{v}_j)\cdot(\vec{r}_i-\vec{r}_j)\over
h_{ij}\left(|\vec{r}_i -\vec{r}_j|^2/h_{ij}^2+\eta^2\right)}{f_i+f_j \over 2
c_{ij}}& if $(\vec{v}_i-\vec{v}_j)\cdot(\vec{r}_i-\vec{r}_j)<0$\cr
	     0& if $(\vec{v}_i-\vec{v}_j)\cdot(\vec{r}_i-\vec{r}_j)\ge0$\cr}.
		\label{muDB}
\end{equation}
Here $f_i$ is the form function for particle $i$ defined by
\begin{equation}
f_i={|{\bf \nabla}\cdot \vec{v}|_i \over |{\bf \nabla}\cdot \vec{v}|_i
+|{\bf \nabla}\times \vec{v}|_i + \eta' c_i/h_i
}, \label{fi}
\end{equation}
where the factor $\eta'\sim 10^{-4}-10^{-5}$ prevents numerical
divergences, $({\bf \nabla}\cdot \vec{v})_i$ is given by equation (\ref{divv}), and
\begin{equation}
({\bf \nabla}\times \vec{v})_i={1 \over \rho_i}\sum_j m_j
	(\vec{v}_i-\vec{v}_j)\times{\bf \nabla}_i W_{ij}. \label{curlv}
\end{equation}
The form function $f_i$ acts as a switch, approaching unity
in regions of strong compression ($|{\bf \nabla}\cdot \vec{v}|_i
>>|{\bf \nabla}\times \vec{v}|_i$) and vanishing in regions of large
vorticity ($|{\bf \nabla}\times \vec{v}|_i >>|{\bf \nabla}\cdot \vec{v}|_i$).  Consequently, this AV has the advantage that it is suppressed in shear layers.  
Note that since $(p_i/\rho_i^2+p_j/\rho_j^2)\approx
2c_{ij}^2/(\gamma\rho_{ij})$, equation~(\ref{piDB})
behaves like equation (\ref{pi}) when $|{\bf
\nabla}\cdot \vec{v}|_i >> |{\bf \nabla}\times \vec{v}|_i$,
provided one rescales the $\alpha$ and $\beta$ in equation (\ref{piDB}) to be a
factor of $\gamma/2$ times the $\alpha$ and $\beta$ in equation (\ref{pi}).

\subsubsection{Thermodynamics}

To complete the description of the fluid, either $u_i$ or $A_i$ is
evolved according to a discretized version of the first law of
thermodynamics.  Although various forms of these evolution equations exist, 
the most commonly used are
\begin{equation}
{{\d} u_i\over \d t}= {1\over 2}\sum_j m_j \left({p_i\over\rho_i^2}+{p_j\over\rho_j^2}+
    \Pi_{ij}\right)\,(\vec{v}_i-\vec{v}_j)\cdot{\bf \nabla}_i
    W_{ij},
	\label{udot}
\end{equation}
and
\begin{equation}
{{\d} A_i\over \d t}={\gamma-1\over 2\rho_i^{\gamma-1}}\,
     \sum_jm_j\,\Pi_{ij}\,\,(\vec{v}_i-\vec{v}_j)\cdot{\bf \nabla}_i
     W_{ij}.
	\label{adot}
\end{equation}
We call equation~(\ref{udot}) the ``energy equation,'' while equation
(\ref{adot}) is the ``entropy equation.''  Which equation one should
integrate depends upon the problem being treated.  Each has its own
advantages and disadvantages. Note that the derivation of
equations~(\ref{udot}) and (\ref{adot}) neglects terms proportional to
the time derivative of $h_i$.  Therefore if we integrate the energy
equation, even in the absence of AV, the total
entropy of the system will not be strictly conserved if the particle
smoothing lengths are allowed to vary in time; if the entropy equation
is used to evolve the system, the total entropy would then be strictly
conserved when $\Pi_{ij}=0$, but not the total energy (Rasio 1991;
Hernquist 1993). For more accurate treatments involving
time-dependent smoothing lengths, see Nelson \& Papaloizou (1994)
and Serna et al. (1996).
The energy equation has the advantage that other thermodynamic processes
such as heating and cooling (Katz et al.\ 1996) and nuclear burning
Garcia-Senz et al.\ 1998) can be incorporated more easily.

\subsubsection{Integration in Time}

The results of SPH simulations involving only hydrodynamic forces 
and gravity do not depend strongly on the actual
time-stepping routine used, as long as the routine remains stable and
accurate.  A simple second-order explicit leap-frog scheme is often
employed. Implicit schemes must be used when other processes such as heating 
and cooling are coupled to the dynamics (Katz et al.\ 1996).
A low order scheme is appropriate for SPH because pressure
gradient forces are subject to numerical noise.  For stability, the
timestep must satisfy a modified Courant condition, with $h_i$ replacing the
usual grid separation.
For accuracy, the timestep must be a small enough fraction of the
 dynamical time.

Among the many possible choices for determining the timestep,
the prescription proposed by Monaghan (1989) is recommended. This sets
\begin{equation}
\Delta t=C_N\,{\rm Min}(\Delta t_1,\Delta t_2), \label{dt}
\end{equation}
where the constant dimensionless Courant number $C_N$ typically
satisfies $0.1\lo C_N \lo 0.8$, and where
\begin{eqnarray}
\Delta t_1 &=&{\rm Min}_i\,(h_i/\dot v_i)^{1/2}, \label{dt1} \\
\Delta t_2 &=&{\rm Min}_i\left(
{h_i \over c_i+k\left(\alpha c_i+\beta {\rm Max_j}|\mu_{ij}|\right)}
\right),  \label{good.dt}
\end{eqnarray}
with $k$ being a constant of order unity.
If the Hernquist \& Katz AV [eq.~(\ref{pi2})] is used, the quantity
Max$_j|\mu_{ij}|$ in equation (\ref{good.dt}) can be replaced by $h_i|{\bf
\nabla}\cdot \vec{v}|_i$ if $({\bf \nabla}\cdot \vec{v})_i<0$, and by
$0$ otherwise.  By accounting for AV-induced diffusion, the $\alpha$ and 
$\beta$ terms in the denominator of equation (\ref{good.dt}) allow 
for a more efficient use of
computational resources than simply using a smaller value of $C_N$.

\subsubsection{Smoothing Lengths and Accuracy}

The size of the smoothing lengths is often chosen such that particles
roughly maintain some predetermined number of neighbors $N_N$.  Typical
values of $N_N$ range from about 20 to 100.  If a particle
interacts with too few neighbors, then the forces on it are sporadic, a
poor approximation to the forces on a true fluid element. In general, one
 finds that, for given physical conditions, 
the noise level in a calculation always decreases when $N_N$ is increased.  

At the other
extreme, large neighbor numbers degrade the resolution by requiring
unreasonably large smoothing lengths.  
However,  higher accuracy is obtained in SPH calculations only when 
{\em both\/}
the number of particles $N$ {\em and\/} the number of neighbors $N_N$ are
increased, with $N$ increasing faster than $N_N$ so that the smoothing
lengths $h_i$ decrease. Otherwise (e.g., if $N$ is increased while maintaining
$N_N$ constant) the SPH method is {\em inconsistent\/}, i.e., it converges to an
unphysical limit. This can be shown easily by deriving the dispersion relation
for sound waves propagating in simple SPH systems (Rasio 1991). The choice of $N_N$ for a given
calculation is therefore dictated by a compromise between an acceptable
level of numerical noise and the desired spatial resolution (which is
$\approx h\propto 1/N_N^{1/d}$ in $d$ dimensions) and level of accuracy.

\subsection{Results of Recent Test Calculations}

The authors and their collaborators have performed a series of
systematic tests to evaluate the effects of spurious transport in SPH
calculations.
These tests are presented in detail in Lombardi et al.\ (1999), 
while here we summarize the main results. Our tests include (i) particle
diffusion measurements, (ii)
shock-tube tests, (iii) numerical viscosity measurements, and
(iv) measurements of the spurious transport of angular momentum due to
AV in differentially rotating, self-gravitating configurations.  
The results are useful for quantifying the
accuracy of the SPH scheme, especially for problems where shear flows
or shocks are present, as well as for problems where true mixing is
relevant.  Other recent
tests of SPH include those by Hernquist \& Katz (1989) and by Steinmetz
\& M\"uller (1993).  

\subsubsection{Particle Diffusion}
 
Many of our tests focus on spurious diffusion, the motion of SPH
particles introduced as an artifact of the numerical scheme.  Often
applications require a careful tracing of the particle positions, and
in these cases it is essential that spurious diffusion be small.  For
example, SPH simulations can be used to establish the degree of
fluid mixing during stellar collisions, which is of
primary importance in determining the subsequent stellar evolution of
the merger remnants (see \S 2.1).  It must be stressed that the amount of
mixing determined by SPH calculations is always an upper limit.
In particular, low-resolution calculations tend to be noisy, and this 
noise can lead to
spurious diffusion of particles, independent of any real physical
mixing of fluid elements.  

We have analyzed spurious diffusion by using SPH particles in 
a box with
periodic boundary conditions to model a stationary fluid of infinite extent. 
For various noise levels (particle velocity dispersions) and neighbor
numbers $N_N$, we measure the rate of diffusion, quantified by the
diffusion coefficient
\begin{equation}
D\equiv\left\langle{{\d}\Delta r^2\over \d t}\right\rangle.
\end{equation}
Here the brackets $\langle\rangle$ denote a time average, and $\Delta
r=(\Delta x^2+\Delta y^2+\Delta z^2)^{1/2}$ is
the total distance traveled by a particle due to spurious diffusion.
Although strong shocks and AV in SPH calculations can lead to 
additional particle mixing (Monaghan
1989), particle diffusion is the dominant contribution to spurious 
mixing in weakly shocked fluids.

Once expressed in terms of 
the number density of SPH particles and the 
sound speed, these diffusion coefficients can therefore be used to
estimate spurious deviations in particle positions in a wide variety 
of applications, including self-gravitating systems.  
For each particle in some large-scale simulation, this spurious 
deviation is estimated simply by 
numerically integrating
\begin{equation}
\Delta r^2\approx \int{ D \d t}. \label{integral}
\end{equation}
The coefficient $D$ in the integrand of equation (\ref{integral})
depends on the particle's velocity deviation from the local flow, the
local number density $n$ of particles, and the local sound speed $c_s$,
so that these quantities need to be monitored for each particle during
the simulation.  Such a scheme was successfully used to estimate
spurious mixing in the context of stellar collisions (Lombardi et al.
1996), where typically (with $N=3\times 10^4$ and $N_N\approx 64$) the
diffusion coefficient was very roughly $D\sim 0.05 c_s n^{-1/3}$.

For sufficiently low noise levels, the diffusion coefficient
essentially vanishes, as the particles simply oscillate around
equilibrium lattice sites.  We say that such a system has
``crystallized.''
For a neighbor number $N_N\approx 64$, a system of SPH particles will 
crystallize if the root mean square velocity dispersion is less than about
3--4\% of the sound speed.
We find that crystallized cubic lattices are unstable against
perturbations, while lattice types with large packing fractions, such as hexagonal
close-packed, are stable.  For this reason it may sometimes be better
to construct initial data by placing particles in an hexagonal
close-packed lattice, rather than in a cubic lattice as is often done.

The diffusion coefficients have been measured using equal-mass
particles.  Sometimes, however, SPH simulations use particles of
unequal mass so that less dense regions can still be highly resolved.
To test the effects of unequal mass particles in a self-gravitating
system, we constructed an equilibrium $n=1.5$ polytrope
(a polytrope is an idealized model for a spherical star, characterized
by a relation of the form $P=\rho^\gamma$ between pressure $P$ and density
$\rho$; the polytropic index $n$ is defined by $\gamma=1+1/n$), using particle
masses which increased with radius in the initial configuration.
Allowing the system to evolve, we observed that the heaviest particles
gradually migrated towards the center of the star, exchanging places
with less massive particles.  For a polytrope modeled with 
$N\approx 1.4\times 10^4$ particles and a neighbor number $N_N\approx 64$, 
the distribution of particle masses is reversed within roughly 
80 dynamical timescales. This is caused by the
interactions among neighboring particles via the smoothing kernel.
These interactions allow energy exchange, and equipartition of energy
then requires the heavier particles to sink into the gravitational
potential well.  Spurious mixing is therefore a more complicated
process in simulations which use unequal mass particles: each particle
has a preferred direction to migrate, and in a dynamical application
this direction can be continually changing.  For simulations in which
fluid mixing is important, equal-mass particles are an appropriate
choice.

\subsubsection{Shock Tube Tests}

The diffusion tests just described are all done in the absence of
shocks and without AV.  To test the AV schemes described in
\S1.2, we turn to a periodic version of the 1-D Riemann shock-tube
problem.  Initially, fluid slabs with constant (and
alternating) density $\rho$ and pressure $p$ are separated by an
infinite number of planar, parallel, and equally spaced interfaces.  We
treat this inherently 1-D problem with both a 1-D and a 3-D SPH code.
The 1-D code is naturally more accurate, and provides a benchmark
against which we can compare the results of our 3-D code.
In both cases, periodic boundary conditions allow us to model 
the infinite number of slabs.

Using various values of $\alpha$ and $\beta$, we performed a number of
such shock tube calculations with our 3-D code, at both Mach numbers ${\cal
M}\approx1.6$ and ${\cal M}\approx 13.2$.
We then compared the
time variation of the internal energy and entropy of the system against
that of the 1-D simulation. Furthermore, since
any motion perpendicular to the bulk fluid flow is spurious, we were also
able to examine spurious mixing in these
simulations.  
We find that all three forms of AV can handle shocks well.  For example, 
with $N=10^4$ and $N_N\approx 64$,
there is better than 2\% agreement with the 1-D code's internal energy
vs.~time curve when ${\cal M}\approx 1.6$, and
agreement at
about the 3\% level when ${\cal M}\approx 13.2$.
We also find that both equations (\ref{pi}) and (\ref{piDB}), as compared to equation (\ref{pi2}), allow less spurious mixing and do somewhat
better at reproducing the 1-D code's results.

Such simulations are a useful and realistic way to calibrate spurious
transport, since the test
problem, which includes shocks and significant fluid motion, has many
of the same properties as real astrophysical problems.  
In fact, the recoil shocks in stellar collisions
do tend to be nearly planar, so that even the 1-D geometry of the shock
fronts is realistic.  The periodic boundary conditions play the role of
gravity in the sense that they prevent the gas from expanding to
infinity.

\subsubsection{Shear Flows}
 
To test the various AV forms in the presence of a shear flow, we impose
the so-called slipping boundary conditions on a periodic box, as is
commonly done in molecular dynamics (see, e.g., Naitoh \& Ono 1976).
The resulting ``stationary Couette flow'' has a velocity field close to
$(v_x,v_y,v_z)=(v_0y/L,0,0)$ and allows us to measure the numerical
viscosity of the particles.  As in the shock tube tests,
we also examine spurious mixing in the direction perpendicular to the fluid flow.  
These shear tests therefore allow us to further investigate the
accuracy of our SPH code as a function of the AV parameters and
scheme.  We find that both the Hernquist \& Katz AV [eq.~(\ref{pi2})] and the
Balsara AV [eq.~(\ref{piDB})] exhibit less viscosity than the
classical AV [eq.~(\ref{pi})].  However, the
classical AV does allow significantly less spurious mixing than the
other forms.  
For all three forms of the AV, increasing $\alpha$ and $\beta$ tends to 
damp out the noise and
consequently decrease spurious mixing, but it also increases the spurious
shear viscosity.

Rotation plays an important role in many hydrodynamic processes.  For
instance, a collision between stars can yield a rapidly and
differentially rotating merger remnant.  Even in the absence of shocks,
AV tends to damp away differential rotation due to the relative
velocity of neighboring particles at slightly different radii, and an
initially differentially rotating system will tend towards rigid
rotation on the viscous dissipation timescale. In systems best modeled
with a perfect fluid, ideally with a viscous timescale $\tau=\infty$,
any such angular momentum transport introduced by the SPH scheme is
spurious.

As a concrete example, we consider an axisymmetric equilibrium
configuration differentially rotating with an angular velocity profile $\Omega(\varpi) \propto
\varpi^{-\lambda}$, where $\varpi$ is the distance from the rotation axis
and $\lambda$ is a constant of order unity.
We then analytically estimate the viscous dissipation timescale
for each of the three AVs
discussed in \S 1.2.  
These analytic estimates are found to closely match numerically measured
values of the timescale.  Both the Hernquist \& Katz AV [eq.~(\ref{pi2})] and
the Balsara AV [eq.~(\ref{piDB})] yield longer viscous timescales than the classical AV [eq.~(\ref{pi})], and hence are better at
maintaining the angular velocity profile.  The Balsara AV clearly does best in this regard, with a viscous timescale roughly $N_N^{1/2}$ times larger than for the classical AV.

When choosing values of AV parameters, 
one must
weigh the relative importance of shocks, shear, and fluid mixing.
For this
reason, it is an application-dependent, somewhat subjective matter to specify ``optimal values''
of $\alpha$ and $\beta$.  We do,
however, roughly delineate the boundaries of the region in parameter
space that gives acceptable results in Lombardi et al.\ (1999).

Our results concerning the various AV forms can be summarized as follows
(see Lombardi et al.\ 1999 for more details).
We find that the AVs defined by equations (\ref{pi}) and (\ref{piDB})
do equally well both in their handling of shocks and in their
controlling of spurious mixing, and do slightly better than equation
(\ref{pi2}).  Furthermore, both equations (\ref{pi2}) and (\ref{piDB}) do
introduce less numerical viscosity than equation (\ref{pi}).
Since equation (\ref{piDB}), Balsara's form of AV, does indeed
significantly decrease the amount of shear viscosity without
sacrificing accuracy in the treatment of shocks, we conclude that it is
an appropriate choice for a broad range of problems.  This is
consistent with the successful use of Balsara's AV reported by Navarro
\& Steinmetz (1997) in their models of rotating
galaxies.

%%%%%%%%%%%%%%%%%%%%%%%%%%%%%%%%%%%%%%%%%%%%%%%%%%%%%%%%%%%%%%%%%%%%%%%%%
\section{SPH Calculations of Stellar Interactions}

The vast majority of recent 3-D calculations of dynamical interactions
between stars have been done using the SPH method. 
These include collisions (Benz \& Hills 1992; Lombardi et al.\ 1996), 
binary coalescence (Davies et al.\ 1994; Rasio \& Shapiro 1995), 
common envelope evolution (Rasio \& Livio 1996; Terman \& Taam 1996),  
accretion flows (Bate \& Bonnell 1997; Theuns et al.\ 1996),
and tidal disruption (Laguna et al.\ 1993).

SPH has many advantages over more
traditional methods of numerical hydrodynamics for these calculations. First,
and perhaps most importantly, the {\em advection\/} of the fluid while the stars
are moving along their initial trajectories is handled very easily by SPH.
For example, in the case of binary coalescence (see \S2.2 below), one often
has to follow the motion of the two stars for several orbital periods before
the final merger occurs. Merely tracking the motion of a star across a large 
3-D grid for many dynamical times can be very challenging when using an Eulerian 
scheme, especially in the presence of a sharp stellar surface (as in the
case of neutron stars, which contain a fairly incompressible fluid).
In addition, other physical processes can be studied much more easily
using a Lagrangian scheme. An example is {\em hydrodynamic mixing\/},
which is a crucial process in the study of certain stellar merger 
processes (see \S2.1).  Since chemical abundances are passively advected 
quantities during a dynamical evolution, the chemical
composition in the final fluid configuration can be determined after
the completion of a calculation simply by noting the original and final
positions of all SPH particles and by assigning particle abundances
according to an initial profile.  
The adaptiveness of the scheme, with particles automatically concentrating
in regions of higher density is also an important advantage, although this
is more crucial in situations involving large density contrasts, such as
in simulations of star formation or galaxy formation.

\subsection{Stellar Collisions}

As a first illustration of the use of SPH for a typical stellar interaction
problem, we summarize in this section recent work
on the numerical calculation of collisions between two main-sequence stars.

\subsubsection{Motivation}

Close dissipative encounters and direct physical collisions between
stars occur frequently in dense star clusters.  The dissipation of
kinetic energy in close stellar encounters can have a direct influence
on the overall dynamical evolution of a cluster.  Observational
evidence for stellar collisions and mergers in globular clusters is
provided by the existence of large numbers of {\em blue stragglers\/}
in these systems. 
These are peculiar main-sequence (hydrogen burning) stars that appear
younger and more massive than all other, normal
 main-sequence stars in the cluster. 
%This has been directly confirmed in some cases
%using spectroscopic observations (Shara et al.\ 1997).

Blue stragglers have long been thought to be formed through the merger
of two lower-mass stars, either in a collision or following binary coalescence  
(see, e.g., the review by Livio 1993). 
Clear indication for a collisional origin of blue stragglers 
has come from observations of globular clusters with the Hubble Space Telescope.
Large numbers of blue stragglers were found to be concentrated in the cores of  
the densest clusters (such as M15 and M30).
%Collisions can happen directly between two single stars only in the cores
%of the densest clusters, but even in somewhat lower-density environments they 
%can also happen indirectly, during resonant interactions involving
%binaries (Sigurdsson \& Phinney 1995). 

Following early numerical work in 2-D (e.g., Shara \& Shaviv 1978),
Benz \& Hills (1987, 1992) performed the first 3-D calculations of
direct collisions between two main sequence stars using SPH.  An
important result of this pioneering study was that collisions
could lead to a thoroughly mixed merger remnant.  The mixing of fresh hydrogen fuel into the core of the
remnant could then reset its nuclear clock, allowing the blue straggler to
burn hydrogen for a full main-sequence lifetime ($t_{MS}\sim10^9\,$yr)
after its formation. 
%This is crucial for the production of blue
%stragglers, which, apart from their higher mass, appear to be
%completely normal main-sequence stars in the clusters.  Even though
%Benz and Hills' collisions at small impact parameters reveal only
%partial mixing, subsequent works generally assumed that all collisional
%remnants were nearly chemically homogeneous, as with normal zero-age
%main sequence stars.

\subsubsection{Recent Results using SPH}

The authors and their collaborators have re-examined collisions between 
main-sequence stars, and, in particular,
the question of mixing during mergers, by performing a set
of numerical hydrodynamic calculations using SPH (Lombardi, Rasio, \& Shapiro 1995, 1996;
Sills et al.\ 1997; Sills \& Lombardi 1997).  
This new work differs from the previous study of Benz \& Hills (1987) 
by adopting more realistic models of globular cluster stars, 
and by performing numerical calculations with increased spatial resolution.
%Benz \& Hills (1987) used an early version of the SPH method and performed
%their calculations with a relatively small number of particles ($N\simeq10^3$). They also
%represented their stars by $n=1.5$ polytropes. 
%However, stars close to the main sequence turnoff point 
%in a cluster have a mass distribution much more centrally 
%concentrated than $n=1.5$ polytropes and are consequently more 
%resistant to being mixed. These high-mass main-sequence stars are 
%the most important for blue straggler production:
%as the cluster evolves via two-body relaxation,
%the more massive stars tend to concentrate
%in the dense cluster core, where the collision rate is highest 
%(see, e.g., Spitzer 1987). In addition,
% collision rates can be increased dramatically
%by the presence of a significant fraction of primordial binaries in the cluster,
%and the more massive stars will preferentially tend to be exchanged into
%such a binary, or collide with another star, following a dynamical interaction 
%between two binaries or between a binary and a single star
%(Sigurdsson \& Phinney 1995). 
%Our more recent SPH calculations were done using 
%$N\sim10^4-10^5$ particles, and the colliding stars are modeled as 
%composite polytropes or with realistic models calculated using a
%stellar structure code.  
%This is particularly important for collisions between two stars of
%different masses, which in general will also have different internal
%structures (Sills \& Lombardi 1997).

\begin{figure}
\caption{
Snapshots of density contours in the orbital plane for a parabolic
collision between two main-sequence stars of masses $M_1$ and $M_2=0.75 M_1$.  The
impact parameter has been chosen such that the corresponding point-mass
orbit would have a pericenter separation $r_p=0.25(R_1+R_2)$, where
$R_1$ and $R_2=0.56R_1$ are the stellar radii.  There are eight density
contours, which are spaced logarithmically and cover four decades down
from the maximum.  The elapsed time in the upper left corner of each
frame is in units of the dynamical timescale $(R_1^3/GM_1)^{1/2}$.
Adapted from Lombardi et al.\ (1996).
} \label{collision}
\end{figure}

Snapshots from one of our recent calculations are shown in Figure~1.
The main new results of these SPH calculations can be summarized as
follows.  Merger remnants produced by parabolic collisions are always
far from chemically homogeneous. In the case of collisions between two
nearly identical stars, the
amount of hydrodynamic mixing during the collision is minimal.  In
fact, the final chemical composition profile is very close to the
initial profile of the parent stars. For two turnoff stars
(i.e., close to hydrogen exhaustion at the center), this means
that the merger remnant is born with very little hydrogen to burn in
its core and, consequently, that the object may not be able to remain
on the main sequence for long.  In the case of a collision between two
stars of different masses, the chemical composition profile of the
merger remnant tends to be more homogeneous, but it remains true that
little hydrogen is injected into its core.

At a qualitative level, these results can be understood very simply in
terms of the requirement of convective (dynamical) stability of the
final hydrostatic equilibrium configurations.  For non-rotating
remnants, convective stability requires 
that the entropic variable $A$ [see eq.~(\ref{eos})]
increase monotonically from the center to the surface (the so-called 
Ledoux criterion). If shock-heating could
be neglected entirely (which is not unreasonable for the low-velocity
collisions occurring in globular clusters), then one could predict the
final composition profile of a merger remnant simply by observing the 
composition and $A$ profiles of the parent stars.
Since fluid elements conserve both their
value of $A$ (in the absence of shocks) and their composition, the final 
composition profile of a remnant
is constructed simply by combining mass shells in order of increasing
$A$, from the center to the outside.
Many features of the results
follow directly.  For example, in the case of a collision between two
identical stars, it is obvious why the composition profile of
the merger remnant remains very similar to that of the parent stars,
since shock-heating is significant only in the outer layers of the
stars, which contain a very small fraction of the total mass.  Furthermore, the
dense, helium-rich material is concentrated in the deep interior
of the parent stars, where shock-heating is negligible, and therefore
it remains concentrated in the deep interior of the final
configuration.

%One may worry that during the later phase of thermal relaxation, where the
%merger remnant contracts back to the main sequence, other transport processes
%such as convection may lead to efficient mixing of the fluid. However, detailed 
%stellar evolution calculations of non-rotating merger remnants 
%(done using a modified
%version of the state-of-the-art Yale stellar evolution code that can handle
%evolution on a thermal time scale; see Sills et al.\ 1997)
%have shown that the chemical composition profiles of the merger
%remnants remain essentially unchanged during this phase. Therefore, we
%conclude that most blue stragglers produced by stellar collisions
%cannot live long as main sequence stars, unless rotationally induced
%mixing during the stellar evolution proves to be important.
%Nevertheless, the resulting blue straggler lifetimes may be consistent
%with observations (Ouellette \& Pritchet 1998).

%%%%%%%%%%%%%%%%%%%%%%%%%%%%%%%%%%%%%%%%%%%%%%%%%%%%%%%%%%%%%%%%%%%%%%%%%%%%%%
\subsection{Coalescing Compact Binaries}

As a second illustration of the use of SPH for calculations of stellar
interactions, we now turn to the dynamical evolution of compact binary
star systems (containing two compact objects---black holes or neutron stars---in
orbit around one another). 

\subsubsection{Motivation}

Coalescing compact binaries are the most promising known sources 
of {\em gravitational radiation\/} that could be
detected by the new generation of laser interferometers now under construction.
These include the Caltech-MIT LIGO (Abramovici et al.~1992; Cutler et al.~1993)
and the European projects VIRGO (Bradaschia et al.\ 1990) and GEO 600 
(Danzmann 1998). In addition to providing a major new confirmation of
Einstein's theory of general relativity, including the first direct
proof of the existence of black holes (Flanagan \& Hughes 1998;
Lipunov et al.\ 1997), the detection of gravitational
waves from coalescing binaries at cosmological distances could provide 
accurate independent measurements of the Hubble constant
and mean density of the universe (Schutz 1986; Chernoff \& Finn 1993; 
Markovi\'c 1993). For an excellent recent review on the detection and sources of 
gravitational radiation, see Thorne (1996).
Coalescing compact binaries are important for other areas of astrophysics as well. 
In particular, many theoretical models of
{\em gamma-ray bursts\/}  have postulated that the energy source for the bursts could
be coalescing compact binaries at cosmological distances 
(Eichler et al.\ 1989; Narayan, Paczy\'nski, \& Piran 1992).
For a discussion of the hydrodynamics of binary coalescence in the context
of gamma-ray burst models, see Ruffert et al.\ (1997) and references therein. 

Here we focus on binaries
containing two neutron stars (hereafter NS), for which the final coalescence is a
purely hydrodynamic process that has been well-studied using SPH
(as well as other methods).

\subsubsection{Hydrodynamics of the Binary Merger Process}

Hydrostatic equilibrium configurations for binary systems 
with sufficiently close components can
become {\em dynamically unstable\/} (Chandrasekhar 1975; Tassoul 1975). 
The physical nature of this instability is common to all 
binary interaction potentials that are sufficiently steeper than $1/r$
(see, e.g., Goldstein 1980, \S 3.6).
It is analogous to the familiar instability of test particles in circular 
orbits sufficiently close to a black hole
(Shapiro \& Teukolsky 1983, \S 12.4). Here, however, it is 
the {\em tidal interaction\/} that is responsible for the 
steepening of the effective interaction potential between the two stars 
and for the destabilization of the circular orbit. The physical properties
of this instability, and its consequences for NS binary coalescence, 
have been studied by the authors and their collaborators, both numerically
using SPH (Rasio \& Shapiro 1994, 1995, hereafter RS; Rasio 1998) and using analytical
perturbation methods (Lai et al. 1993, 1994, hereafter LRS; Lombardi et al.\ 1997).
%The tidal interaction exists of course already in Newtonian gravity and
%the instability is therefore present even in the absence of relativistic
%effects. For sufficiently compact binaries, however, the combined effects
%of relativity and hydrodynamics lead to an even stronger tendency towards
%dynamical instability (see, e.g., Lombardi et al.\ 1997 and references
%therein). 

%The stability properties of close NS binaries depend sensitively on the NS EOS.
%Close binaries containing
%NS with stiff EOS (adiabatic exponent $\gamma\go2$ if $P=K\rho^\gamma$, where
%$P$ is pressure and $\rho$ is density)
%are particularly susceptible to a dynamical instability. This is because tidal
%effects are stronger for stars containing a less compressible fluid (i.e., for
%larger $\gamma$).
%As the dynamical stability limit is approached, the secular orbital
%decay driven by gravitational wave emission can be dramatically accelerated.
%The two stars then plunge rapidly toward each other, and merge together 
%into a single object in just a few rotation periods. 

The instability was first identified by RS using SPH calculations 
where 
the evolution of binary equilibrium configurations was calculated
for two identical polytropes with $\gamma=2$. 
It was found that when $r\lo3R$ ($r$ is the binary separation and $R$
the radius of an unperturbed NS),
the orbit becomes unstable to radial perturbations and the two stars
undergo rapid coalescence. 
For $r\go3R$, the system could be evolved dynamically
for many orbital periods without showing any sign of orbital evolution
(in the absence of dissipation). 
Many of the results derived in RS and LRS concerning the 
stability properties of NS binaries have 
been confirmed recently in completely independent work by 
New \& Tohline (1997) and by Zhuge, Centrella, \& McMillan (1996).
New \& Tohline (1997) used completely different numerical methods (a combination of
a 3-D Self-Consistent Field code for constructing equilibrium configurations
and a grid-based Eulerian code for following the dynamical evolution of the
binaries), while Zhuge et al.\ (1996) used SPH. 

\subsubsection{Typical SPH Results}

For simplicity, we describe here the dynamical evolution of an unstable, 
initially synchronized binary containing two identical stars.
Typical SPH results for this case are shown in Figure~2. 
During the initial, linear stage of the instability, 
the two stars approach each other and come 
into contact after about one orbital revolution. In the corotating 
frame of the binary, the relative velocity 
remains very subsonic, so that the evolution is adiabatic at this stage.
This is in sharp contrast to the case of a head-on collision between
two stars on a free-fall, radial orbit, where
shocks can be very important for the dynamics.
Here the stars are constantly being held back by a (slowly receding)
centrifugal barrier, and the merging, although dynamical, is much more gentle. 

\begin{figure}
\caption{
Evolution of an unstable binary containing two identical neutron stars.
The stars are modeled as polytropes with $\gamma=3$, corresponding to a stiff 
nuclear equation of state. Projections of all SPH particles onto the orbital plane of the
binary are shown at different times (the orbital motion is in the counterclockwise
direction). Units are such that $G=M=R=1$, where $G$ is the gravitational constant and
$R$ and $M$ are the radius and mass of an unperturbed (spherical) star.
The initial orbital period is $P_{orb}\simeq 24$ in these units. 
Adapted from Rasio \& Shapiro (1994).
} \label{coal}
\end{figure}

After typically 2--3 orbital revolutions the innermost cores of the 
two stars have merged and the system resembles a single, very elongated ellipsoid.
At this point a secondary instability occurs: {\em mass shedding\/} 
sets in rather abruptly. Material is ejected through the outer Lagrange
points of the effective potential and spirals out rapidly.
In the final stage, the spiral arms widen and merge together.
The relative radial velocities of neighboring arms as they merge are supersonic,
leading to some shock-heating and dissipation.
As a result, a hot, nearly axisymmetric rotating halo forms around the central
dense core. 
The halo contains about 20\% of the total mass and the rotation profile
is close to a pseudo-barotrope (Tassoul 1978), with the angular velocity 
decreasing as a power-law $\Omega\propto \varpi^{-\nu}$ where $\nu\lo2$ and $\varpi$ 
is the distance to the rotation axis. The core is rotating uniformly near 
breakup speed and contains about 80\% of the mass still in a cold, degenerate state.
If the initial NS had masses close to $1.4\,M_\odot$, then most recent stiff EOS
would predict that the final merged configuration is still stable 
and will not immediately collapse to a black hole, although it might ultimately
collapse to a black hole as it continues to lose angular momentum.
The emission of gravitational radiation
during dynamical coalescence can be calculated perturbatively
using the quadrupole approximation (RS).
Both the frequency and amplitude of the emission peak somewhere during
the final dynamical coalescence, typically just before the onset of
mass shedding. Immediately after the peak, the amplitude drops abruptly
as the system evolves towards a more axially symmetric state.

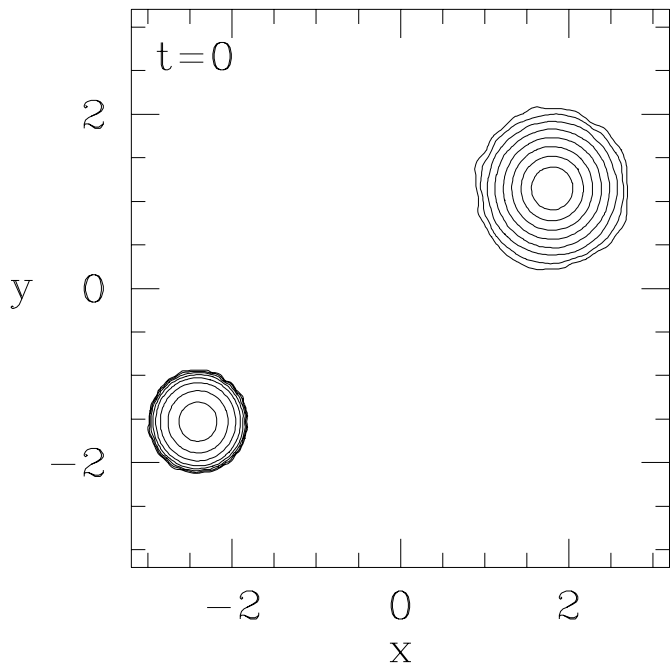
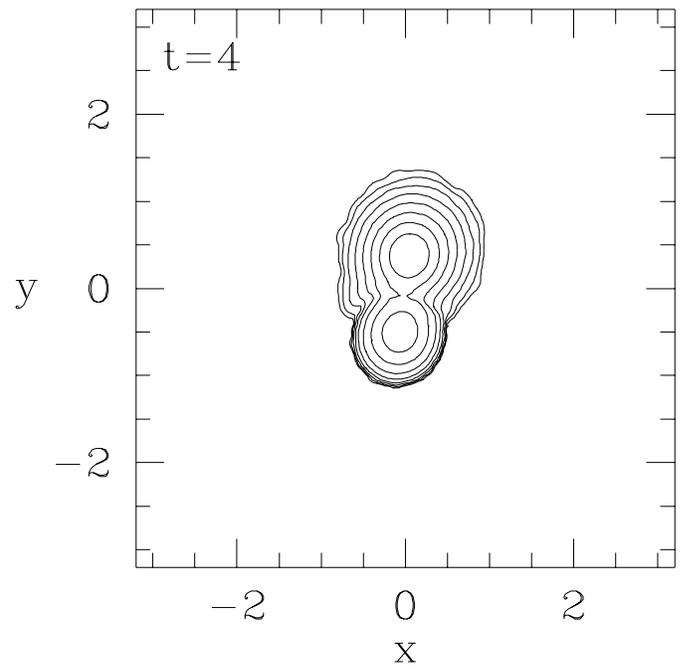
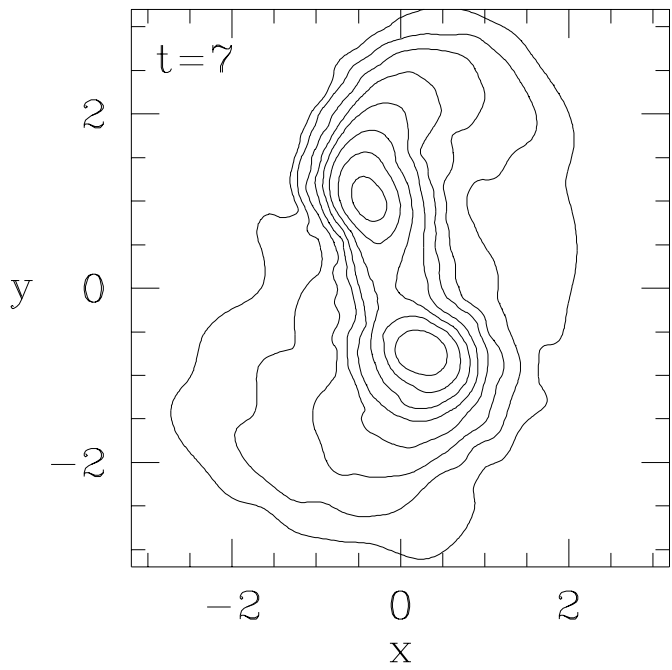
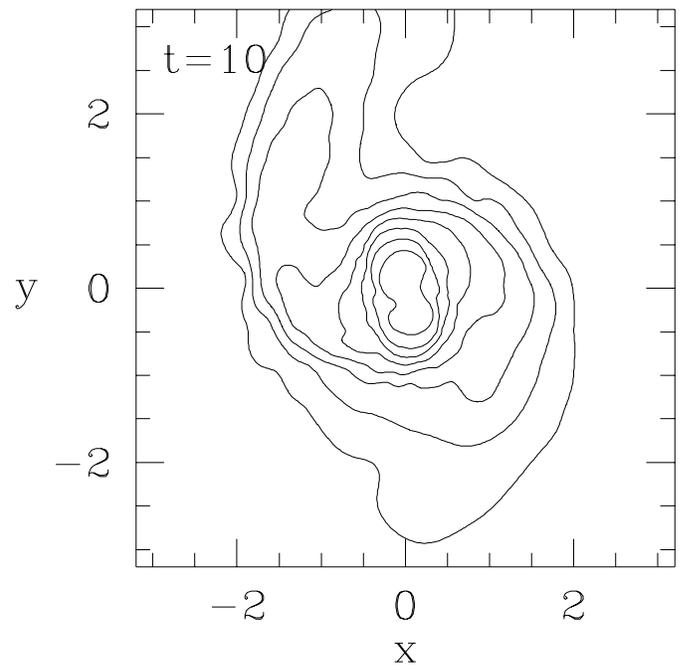
\end{document}